\documentclass[aps,prl,twocolumn,superscriptaddress]{revtex4-1}

\setcitestyle{super}

\usepackage{graphicx,siunitx,braket,xspace,amssymb,dsfont,bm,natbib}
\usepackage{todonotes}
\newcommand{\qS}[0]{\ensuremath{\Ket{\mathrm{S}}}\xspace}
\newcommand{\qT}[0]{\ensuremath{\Ket{\mathrm{T_0}}}\xspace}
\newcommand{\qDU}[0]{\ensuremath{\Ket{\downarrow\uparrow}}\xspace}
\newcommand{\qUD}[0]{\ensuremath{\Ket{\uparrow\downarrow}}\xspace}
\newcommand{\qZ}[0]{\ensuremath{\Ket{0}}\xspace}
\newcommand{\qO}[0]{\ensuremath{\Ket{1}}\xspace}

\newcommand{\rfig}[1]{Fig.\,\ref{#1}}

\newcommand{\rtab}[1]{Tab.\,\ref{#1}}
\newcommand{\citec}[1]{\cite{#1}}

\newcommand{\Tte}[0]{\ensuremath{T_2^{\mathrm{echo}}}\xspace}
\newcommand{\Tts}[0]{\ensuremath{T_2^*}\xspace}
\newcommand{\mc}[1]{\ensuremath{\mathcal{#1}}}
\newcommand{\mr}[1]{\ensuremath{\mathrm{#1}}}
\newcommand{\je}[0]{\ensuremath{J(\epsilon)}\xspace}
\newcommand{\jz}[0]{\ensuremath{J_0}\xspace}
\newcommand{\dbz}[0]{\ensuremath{\mr{\Delta}B_z}\xspace}
\newcommand{\eps}[0]{\ensuremath{\epsilon}\xspace}
\newcommand{\epsz}[0]{\ensuremath{\epsilon_0}\xspace}
\newcommand{\F}[0]{\ensuremath{\mc{F}}\xspace}
\newcommand{\Le}[0]{\ensuremath{\mc{L}}\xspace}

\newcommand{\stz}[0]{$\mathrm{S\mbox{-}T_0}$\xspace}
\newcommand{\figlabel}[1]{\textbf{#1}}

\newcommand{\citenumber}[1]{\hspace{-1 ex} \citenum{#1}} 

\begin{document}

\title{Feedback-tuned noise-resilient gates for encoded spin qubits}

\author{Pascal Cerfontaine\textsuperscript{*}}
\author{Tim Botzem\textsuperscript{*}}
\author{Simon Sebastian Humpohl}
\affiliation{JARA-Institute for Quantum Information, RWTH Aachen University, D-52074 Aachen, Germany}
\author{Dieter Schuh}
\affiliation{Institut f\"ur Experimentelle und Angewandte Physik, Universit\"at Regensburg, D-93040 Regensburg, Germany\\\textsuperscript{*}These authors contributed equally to this work}
\author{Dominique Bougeard}
\affiliation{Institut f\"ur Experimentelle und Angewandte Physik, Universit\"at Regensburg, D-93040 Regensburg, Germany\\\textsuperscript{*}These authors contributed equally to this work}
\author{Hendrik Bluhm}
\affiliation{JARA-Institute for Quantum Information, RWTH Aachen University, D-52074 Aachen, Germany}
\date{\today}


\maketitle
\textbf{Two level quantum mechanical systems like spin 1/2 particles lend themselves as a natural qubit implementation \cite{DiVincenzo1998}. However, encoding a single qubit in several spins reduces the resources necessary for qubit control and can protect from decoherence channels \cite{Levy2002}. While several varieties of such encoded spin qubits have been implemented, accurate control remains challenging, and leakage out of the subspace of valid qubit states is a potential issue. Here, we realize high-fidelity single qubit operations for a qubit encoded in two electron spins in GaAs quantum dots by iterative tuning of the all-electrical control pulses. Using randomized benchmarking \cite{Magesan2011}, we find an average gate fidelity of $\F = \mathbf{(98.5 \pm 0.1)\,\%}$ and determine the leakage rate between the computational subspace and other states to $\Le = \mathbf{(0.4\pm0.1)\,\%}$ \cite{Epstein2014, Wallman2014}. These results also demonstrate that high fidelity gates can be realized even in the presence of nuclear spins as in III-V semiconductors. }

Spins captured in semiconductor nanostructures provide a solid-state approach to quantum computation which leverages current semiconductor production technology for device fabrication. While the two spin states of an isolated electron form a natural qubit, the microwave signals required for manipulation impose certain drawbacks. Hence, all-electrical control is an attractive alternative that can be achieved by encoding a qubit in multi-electron states. Most of the basic operations required for quantum computation have already been demonstrated experimentally for qubits using one \cite{Elzerman2004, Nowack2008}, two \cite{Petta2005, Foletti2009, Shulman2012} and three spins \cite{Gaudreau2011, Medford2013, Kim2015}.

A key requirement for quantum computation is that qubit manipulations, so-called gates, are highly accurate. Corresponding figures of merit are the gate error rate $r$ or the gate fidelity $\F \propto 1-r$. Fidelities well above \SI{99}{\%} are expected to be needed for scalable quantum computing \cite{Fowler2009}. 

Recent works have demonstrated \SI{99}{\%} \cite{Kawakami2016}, \SI{99.6}{\%} \cite{Veldhorst2014} and in one case \SI{99.95}{\%} \cite{Muhonen2015} using AC-controlled single-spin qubits in Si-based systems. Furthermore, fidelities of $93-\SI{96}{\%}$ have been demonstrated for a spin-charge hybrid qubit in Si \cite{Kim2015} and about \SI{96}{\%} for a single spin in GaAs \cite{Yoneda2014}.

However, for purely spin-encoded multi-electron qubits recent theoretical gate constructions \cite{Wang2014, Khodjasteh2012} have not yet been complemented by a systematic experimental effort to achieve high fidelities. Doing so entails a number of difficulties: The large pulse amplitudes required for fast control are prone to systematic errors and render standard Rabi driving inappropriate. Furthermore, nonlinearities in the electric control and a dependence of the noise sensitivity on the qubit control signal make optimal gate constructions nontrivial. In addition to charge noise present in all spin qubit variants \cite{Dial2013}, magnetic field fluctuations from nuclear spins are a major challenge in GaAs \cite{Reilly2008}.

In this work, we develop high-fidelity baseband control for a two-spin qubit in a gate-defined GaAs double quantum dot encoded in the subspace with zero net spin ($s_z = 0$). We address the aforementioned difficulties by numerically tailoring control pulses to our experiment \citec{Cerfontaine2014}. Remaining inaccuracies in these pulses are removed by a control loop termed GAMBIT (Gate Adjustment by Iterative Tomography) which allows the iterative tune-up of gates using feedback  \cite{Cerfontaine2014}. In contrast to control loops based on randomized benchmarking (RB)\cite{Magesan2011}, which have already been applied to superconducting qubits \cite{Egger2014, Kelly2014}, GAMBIT extracts tomographic information to improve convergence. Additionally, we optimize about half an order of magnitude more parameters than related work on superconducting qubits \cite{Kelly2014} to fully leverage the degrees of freedom provided by our hardware.

Using RB, we demonstrate fidelities of \SI{98.5}{\%}. We find the fidelity to be limited more by charge noise than by nuclear spin fluctuations, which are often considered a major hurdle for GaAs qubits. The relatively weak effect of nuclear spins is due to a noise-canceling character of our optimized gates. In addition, we use RB to characterize leakage \cite{Epstein2014, Wallman2014} out of and back into the $s_z = 0$ subspace, an important figure of merit for any encoded qubit. 

\begin{figure}
	\centering
	\includegraphics[width=6.93cm]{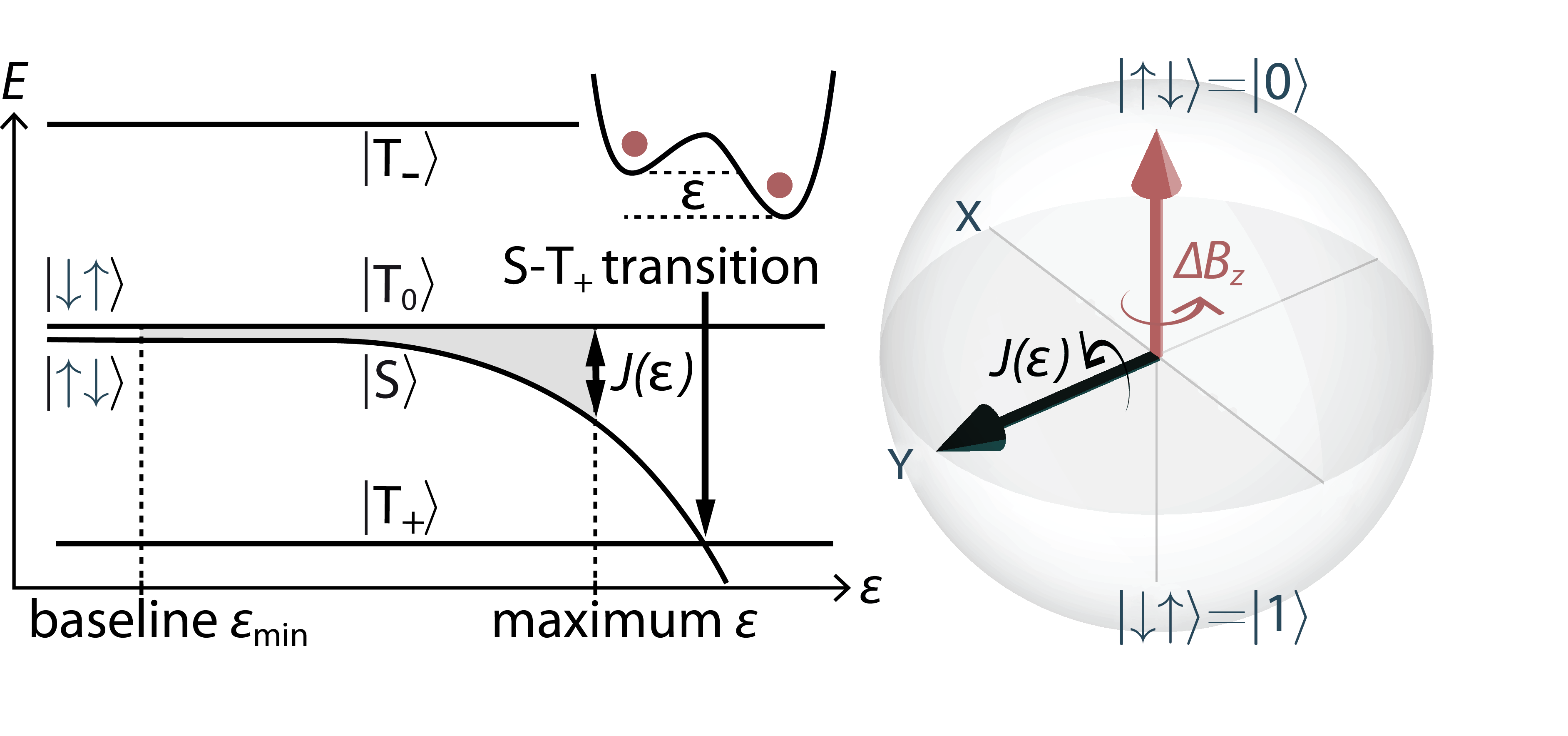}
	\caption{\textbf{$\mathbf{S-T_0}$ qubit energy diagram and Bloch sphere.} The eigenenergies change as a function of detuning \eps, which is used to control the exchange coupling \je. The \eps pulses presented in this work start and finish at a baseline and pulse to higher amplitudes for short periods. The maximum amplitude is constrained to below the $\mathrm{S\mbox{-}T_+}$ anticrossing at large \eps. For ease of understanding we choose the convention that \je points along the \textit{y}-axis of the Bloch sphere (see supplementary information). For low \eps amplitudes, the qubit rotates about \dbz, the \textit{z}-axis of the Bloch sphere. Large amplitude \eps pulses rotate the qubit about the \textit{y}-axis and thus enable arbitrary single-qubit gates.}
\label{fig:conv}	
\end{figure}

\begin{figure*}
	\centering
	\includegraphics[trim=0 0.0cm 0.3cm 0, clip, width=12.08cm]{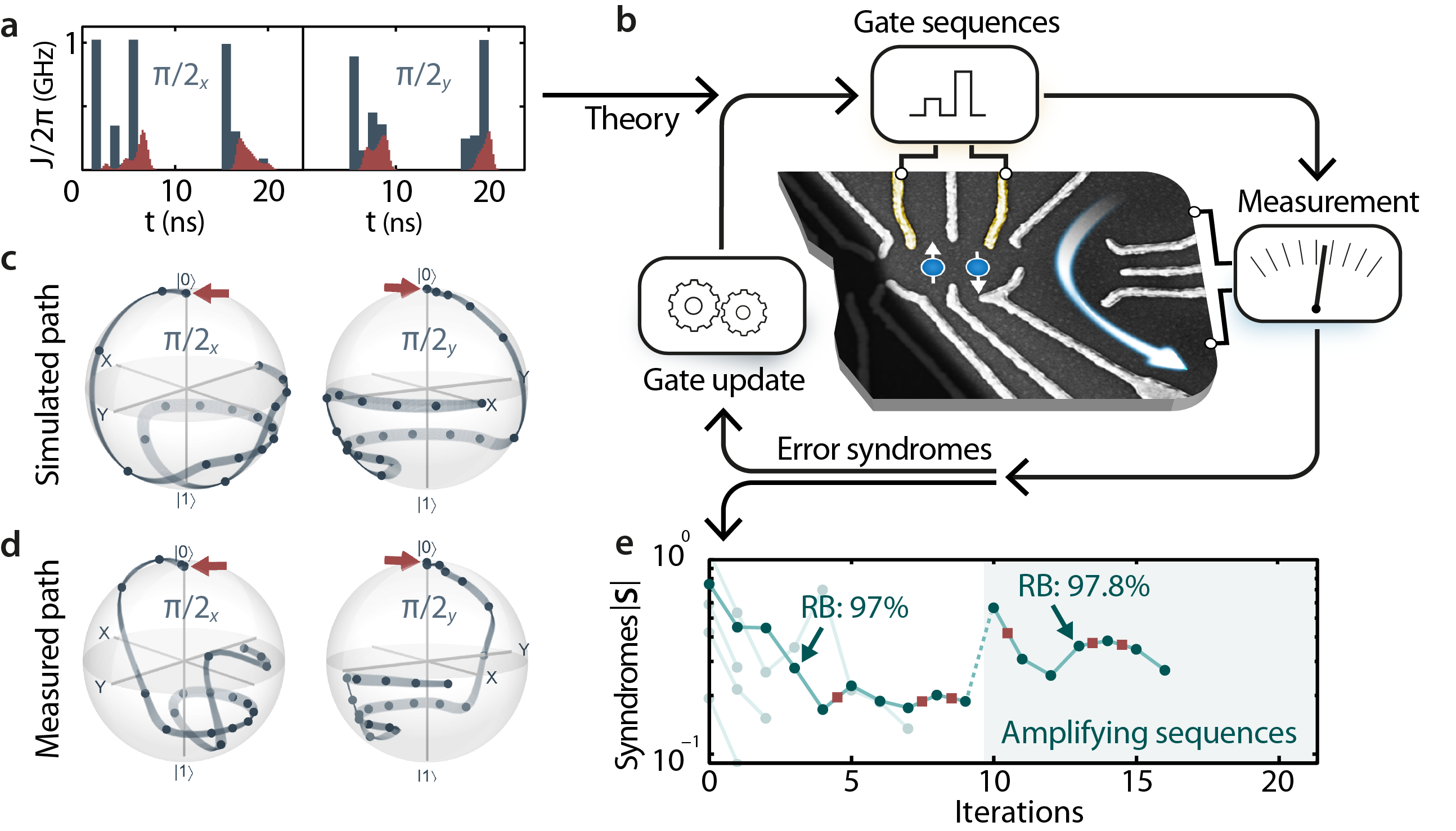}
	\caption{\textbf{Gate adjustment by iterative tomography.} \figlabel{a} Numerical pulse optimization based on a realistic but inaccurate qubit model provides initial optimal control pulses (blue) for $\pi/2_x$ and $\pi/2_y$ gates (\SI{24}{ns} long gates shown here). According to the model, the pulses shown in red are actually seen by the qubit. 
\figlabel{b} Next, these pulses are optimized on the experiment using GAMBIT. 8 error syndromes $\tilde{S_i}$ are extracted in each iteration by applying the gate sequences from \rtab{tab:bs}. In order to remove gate errors, the $\tilde{S_i}$ are minimized by adjusting the pulse segments' amplitudes $\eps_j^g$. After a few iterations, the predicted Bloch sphere trajectories \figlabel{c} can be reproduced in the experiment \figlabel{d} as confirmed by self-consistent state tomography \cite{Takahashi2013}. A major portion of the remaining deviation can be attributed to concatenation errors with the measurement pulses, specifically when the states following large $J$ pulses are determined.
\figlabel{e} Typically, GAMBIT converges within 5 iterations and can recover from charge rearrangements in the quantum dot (indicated by red squares, see supplementary information). For a given noise level, better gates can be achieved by using modified gate sequences which amplify gate errors and lead to larger $\tilde{S_i}$ for the same errors. In this specific run, randomized benchmarking\cite{Magesan2011} (RB) confirms that \F of the gate set was first improved to \SI{97}{\%} and then to \SI{97.8}{\%} by using amplifying gate sequences. Other optimization runs are shown in light blue for comparison.
}
\label{fig:main_1}	
\end{figure*}

Our \stz spin qubit \cite{Levy2002} (see methods and \rfig{fig:main_1}b) can be described by the Hamiltonian $H = \frac{\hbar \je}{2} \sigma_x + \frac{\hbar \dbz}{2} \sigma_z$ in the $\{\qUD = \qZ, \qDU = \qO\}$ basis, where arrows denote electron spin up and down states. \je denotes the exchange splitting between the singlet $\qS = (\qUD-\qDU)/\sqrt{2}$ and $s_z=0$ triplet state $\qT = (\qUD+\qDU)/\sqrt{2}$, while \dbz is the magnetic field gradient across both dots from different nuclear spin polarizations \cite{Petta2005}. The remaining triplet states, $\ket{\mathrm{T_+}} = \ket{\uparrow\uparrow}$ and $\ket{\mathrm{T_-}} = \ket{\downarrow\downarrow}$, represent undesirable leakage states. \je is manipulated by the detuning \eps, the potential difference between both dots. We use standard state initialization and readout (see methods). For single qubit operations, \eps is pulsed on a nanosecond timescale using an arbitrary waveform generator (AWG) whereas \dbz is typically stabilized at $2 \pi \cdot \SI{61.6 \pm 2.5}{MHz}$ by dynamic nuclear polarization \cite{Bluhm2010}. The resulting dynamics are illustrated in \rfig{fig:conv}, using the convention that $\je$ points along the Bloch sphere's \textit{y}-axis for ease of understanding (see supplementary information). A perfect gate implementation is hindered by decoherence due to fluctuations in both \dbz and \eps. Moreover, an imperfectly known nonlinear transfer function \je and finite bandwidth of the voltage pulses can be the source of systematic errors whose elimination requires careful calibration. In our simulations we use the experimentally motivated model $\je = J_0 \exp{(\eps/\eps_0)}$\cite{Dial2013}. 

To experimentally implement accurate single qubit $\pi/2$ rotations around the $x$- and $y$-axis (denoted by $\pi/2_x$ and $\pi/2_y$), we use a control loop adapted from Ref. \citenumber{Cerfontaine2014} (\rfig{fig:main_1}a-b). To obtain a reasonably accurate system model, we measure the step response of our electrical setup, \jz, \epsz and \dbz as well as the coherence properties of the qubit (see supplementary information). We then use this model to numerically optimize pulses consisting of $N_\mathrm{seg}$ piece-wise constant nominal detuning values $\eps_j, j = 1 \dots N_\mathrm{seg}$ to be programmed into the AWG with a segment duration of \SI{1}{ns}. The last four to five segments are set to the same baseline level $\eps_{\min}$ for all gates to minimize  errors arising from pulse transients of previous pulses. We choose $\eps_{\min}$ such that $J(\eps_{\min}) \ll \dbz$. Typical optimized pulse profiles $\eps_j^g, j = 1 \dots N_\mathrm{seg}$ for two gates $g = \pi/2_x$ and $g = \pi/2_y$ are shown in \rfig{fig:main_1}a. 

Since our control model does not capture all effects to sufficient accuracy, these pulses need to be refined using experimental feedback. Hence, error information about the gate set is extracted in every iteration of our control loop. Standard quantum process tomography cannot be applied to extract this information as it requires well-calibrated gates, which are not available before completion of the control loop. We solve this bootstrap problem with a self-consistent method that extracts 8 error syndromes $S_i, i = 1 \ldots 8$ in each iteration \cite{Cerfontaine2014}. $S_i, i = 1 \dots 6$ is primarily related to over-rotation and off-axis errors while $S_i, i \in \{7, 8\}$ are proxies for decoherence. A syndrome $S_i$ is measured by preparing \qZ, applying the corresponding sequence $U_i$ of gates from \rtab{tab:bs}, and determining the probability $p(\qZ)$ of obtaining the state \qZ by measuring the sequence $10^3 \ldots 10^4$ times. For perfect gates, the first six syndromes\cite{Dobrovitski2010} should yield $p(\qZ) = 0.5$, corresponding to $S_i = \left<\sigma_z\right> = 0$. The last two syndromes should yield $p(\qZ) = 0$ ($S_i = -1$). Deviations of $S_i$ from expectation indicate decoherence and systematic errors in the gate set. To make our method insensitive to state preparation and measurement (SPAM) errors, we also prepare and read out a completely mixed state with measurement result $S_{\mathrm{M}}$, and a triplet state \qT, which yields the measurement result $S_{\mathrm{T}}$ after correcting for the approximate contrast loss of the triplet preparation (see supplementary information). GAMBIT then minimizes the modified error syndromes $\tilde{S_i} = |S_i - S_{\mathrm{M}}|$ for $i = 1 \ldots 6$ and $\tilde{S_i} = |S_i - S_{\mathrm{T}}|$ for $i \in \{7, 8 \}$.

For swift convergence, we start the control loop with pulses $\eps_j^g$ which theoretically implement the desired operations perfectly with minimal decoherence. First, GAMBIT scales these pulses by \SI{\pm 20}{\%} in \SI{4}{\%} increments and measures which scaling achieves the lowest $\tilde{S_i}$. GAMBIT then optimizes the best pulses by minimizing $\tilde{S_i}$ with the Levenberg-Marquardt algorithm (LMA). In each LMA iteration, we use finite differences to experimentally estimate derivatives $d\tilde{S_i}/d\eps_j^g$, which are subsequently used to calculate updated pulse amplitudes $\eps_j^g$.

\begin{table}[b]
\small
  \centering
  \caption{\textbf{Tomographic gate sequences.} To first order, the outcome of the measurement $\mr{Tr}(\sigma_z U_i \qZ\!\Bra{0} U_i^\dagger) = S_i$ depends linearly on the gates' rotation-angle errors $2\phi$ ($2\chi$), the axis-errors $n_y, n_z$ ($v_x, v_z$) and decoherence $d_x$ ($d_y$) of the $\pi/2_x$-gate ($\pi/2_y$-gate). Parametrization defined as in Refs. \citenumber{Cerfontaine2014} and \citenumber{Dobrovitski2010} (see supplementary information).}
  	\begin{ruledtabular}
    \begin{tabular}{rrcl}
    Sequences $U_i$ (right to left) & Parametrization & & $S_i$ \\
    \hline
    $\pi/2_x$ & $-2\phi$  & = &$S_1$\\
    $\pi/2_y$ & $-2\chi$  & = &$S_2$\\
    $\pi/2_y \circ \pi/2_x$ & $-n_y - n_z - v_x - v_z$  & = &$S_3$\\
    $\pi/2_x \circ \pi/2_y$ & $-n_y + n_z - v_x + v_z$  & = &$S_4$\\
    $\pi/2_x \circ \pi/2_x \circ \pi/2_x \circ \pi/2_y$ & $n_y + n_z + v_x - v_z$  & = &$S_5$\\
    $\pi/2_x \circ \pi/2_y \circ \pi/2_y \circ \pi/2_y$ & $n_y - n_z + v_x + v_z$  & = &$S_6$\\
    $\pi/2_x \circ \pi/2_x$ & $d_x$ & = & $S_7$\\
    $\pi/2_y \circ \pi/2_y$ & $d_y$ & = & $S_8$\\       
    \end{tabular}%
    \end{ruledtabular}    
  \label{tab:bs}%
\end{table}%

\begin{figure*}
	\centering
	\includegraphics[width=12.32cm]{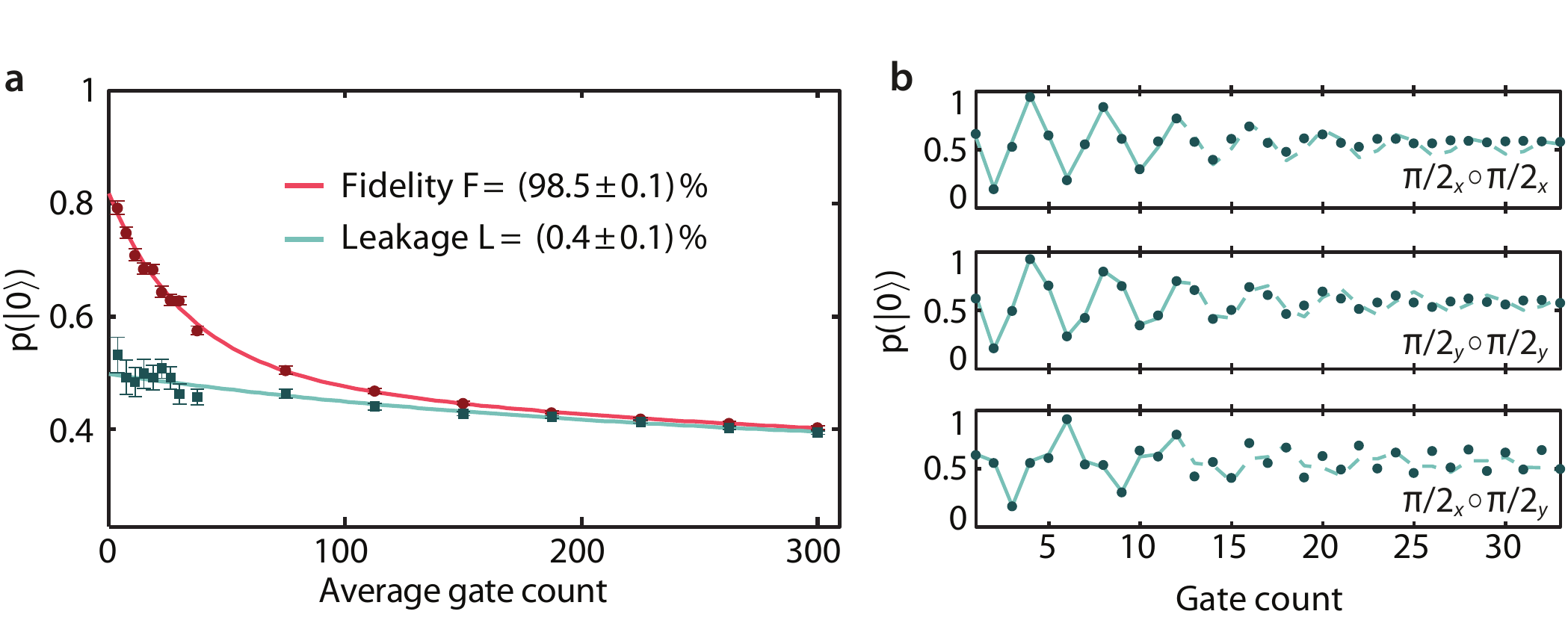}
	\caption{\textbf{Characterization of optimized gate sets.} \figlabel{a} The overall fidelity of a gate set consisting of \SI{30}{ns} long $\pi/2_x$ and $\pi/2_y$ gates is determined using RB (red). Each red data point is an average over 50 randomly chosen sequences of the respective length. In order to determine gate leakage we supplement the standard protocol (red) by a variant which omits the last inversion pulse (blue) \cite{Wallman2014}. Fitting both curves simultaneously with a double (red) \cite{Epstein2014} and a single exponential (blue) yields $\F = \SI{98.5 \pm 0.1}{\%}$ and a leakage rate $\Le = \SI{0.4 \pm 0.1}{\%}$. \figlabel{b} In order to determine systematic errors, we measure multiple repetitions of $\pi/2_x$, $\pi/2_y$ and $\pi/2_x \circ \pi/2_y$, using gates obtained in another GAMBIT run. The fit shown in blue estimates systematic errors of \SI{0.7}{\%}. Since we use a depolarizing channel for a simplified decoherence model, we only fit the first 12 data points. Afterwards other decoherence processes like \Tts effects become dominant.
}
\label{fig:main_2}	
\end{figure*}

Pulses with $N_\mathrm{seg} \ge 24$ lead to reliable convergence, typically within 5 iterations (\rfig{fig:main_1}e). To demonstrate that our approach is reproducible for different initial gates, \SI{24}{ns} gates were used in \rfig{fig:main_1} while the experiments in \rfig{fig:main_2} were performed using \SI{30}{ns} gates. GAMBIT usually only adjusts those segments $\eps_j^g$ which are not at the baseline, resulting in 14 (24) free parameters for the \SI{30}{ns} $\pi/2_x$ ($\pi/2_y$) gate shown in \rfig{fig:main_2}. When convergence eventually slows, we apply the sequences from \rtab{tab:bs} multiple times to amplify certain systematic gate errors (see supplementary information). Thus, further improvement (shaded region in \rfig{fig:main_1}e) is possible without increasing the averaging time per iteration. 

Unfortunately, frequent charge rearrangements in our sample lead to a deterioration of optimized gates within minutes to hours. As a remedy we run GAMBIT again, resulting in slightly different gates than before. For this reason, the experiments in \rfig{fig:main_2}a and \rfig{fig:main_2}b were performed with gates from different GAMBIT runs.

To visualize the experimental gates, we perform self-consistent quantum state tomography (QST) \cite{Takahashi2013} and extract state information after each segment $\eps_j^g$. As seen in \rfig{fig:main_1}c-d, the qubit state trajectories for model and experiment closely resemble each other, indicating that the GAMBIT-tuned pulses remain close to the optimum found in simulations.

In order to rigorously determine \F, we apply RB after completion of GAMBIT. In RB, \F is obtained by applying sequences of randomly chosen Clifford gates, composed of $\pi/2_x$ and $\pi/2_y$ gates, to the initial state \qZ. The last Clifford operation of each sequence is chosen such that \qZ would be recovered if the gates were perfect \cite{Magesan2011}. For imperfect gates, the return probability $p(\qZ)$ decays as a function of sequence length and the decay rate indicates the average error per gate.

We find that the measured decay curve shown in red in \rfig{fig:main_2} is best fitted by a double exponential, with the slow time constant describing the decay beyond $\sim 100$ gates. In some cases, such a deviation from a single exponential decay can arise from non-Markovian noise \cite{Veldhorst2014} or inhomogeneous broadening of the control \cite{Ryan2008}. In our case, we attribute the second decay rate to gate leakage out of the computational subspace \cite{Epstein2014} into $\ket{\mathrm{T_+}} = \ket{\uparrow\uparrow}$. 

To confirm this hypothesis, we apply an extended RB protocol which omits the last Clifford from each RB sequence \cite{Wallman2014}. Without leakage, averaging over many randomly chosen sequences should yield $p(\qZ) =  \SI{50}{\%}$. However, for nonzero leakage we expect a single exponential decay of $p(\qZ)$ as a function of increasing sequence length since the additional leakage states have the same readout signature as \qO (see methods). We indeed find such a decay law, indicated in blue in \rfig{fig:main_2}a. A joint fit of the standard (red) and leakage detection (blue) RB data yields $\F = \SI{98.5+-0.1}{\%}$ and a gate leakage rate $\Le = \SI{0.4+-0.1}{\%}$ (the sum of leakage out of and back into the computational subspace \cite{Wallman2014}). Both fitted decay curves asymptotically approach $p(\qZ) = \SI{0.36}{}$ for long gate sequences, close to $\frac{1}{3}$ as expected for a single leakage state \cite{Epstein2014, Wallman2014}. Since our pulses operate close to the $\mathrm{S}-\mathrm{T_+}$ transition while $\ket{\mathrm{T_-}}$ is far away in energy, leakage should predominantly occur into the $\ket{\mathrm{T_+}}$ level.

As RB does not reveal whether our gates are limited by systematic errors or decoherence, we perform an independent test by measuring repetitions of $\pi/2_x$, $\pi/2_y$ and $\pi/2_x \circ \pi/2_y$ as shown in \rfig{fig:main_2}b. By fitting this data (see supplementary information) we retrieve $\F_{\mathrm{sys}} = \SI{99.3}{\%}$, excluding decoherence. RB yields $\F = \SI{98.1+-0.2}{\%}$ for this gate set, indicating that decoherence and systematic errors contribute roughly equally. Note that even for a bare $T_2^{\star}$ time of less than \SI{100}{ns} along either $J$ or $\Delta B_z$, the decay time for both gates exceeds \SI{500}{ns}. This behavior is expected since the numerically optimized gates exhibit a reduced sensitivity to quasistatic noise sources\cite{Cerfontaine2014}, which contribute significantly to decoherence. In addition, the numerical optimization minimizes the use of large $J$ to increase the resilience to slow and fast charge noise.

We previously predicted fidelities approaching \SI{99.9}{\%} for GaAs based \stz qubits \cite{Cerfontaine2014} with the best reported noise levels \cite{Dial2013, Bluhm2010}. To determine why our gates perform worse, we measure \Tts and \Tte for both exchange and hyperfine driven oscillations. We find that our sample suffers from much larger charge noise than reported in Ref. \citenumber{Dial2013}, which shows up as a shorter $\Tte = \SI{183}{ns}$ for exchange oscillations at $\je = 2 \pi \cdot\SI{61}{MHz}$, compared to $\Tte \approx \SI{7.5}{\micro s}$ at comparable charge noise sensitivity $dJ/d \eps \approx 2 \pi \cdot \SI{150}{MHz/mV}$ \cite{Dial2013}. Using a noise model based on these measurements, we predict fidelities of \SI{98.6}{\%} and \SI{99.0}{\%} for the numerically optimized gates used as a starting point for GAMBIT (see supplementary information). These are close to the experimental value of $\SI{98.5+-0.1}{\%}$ which supports the validity of our noise model and the predictions of Ref. \citenumber{Cerfontaine2014} that a substantial improvement is possible with previously measured lower charge noise levels. Enhanced suppression of hyperfine fluctuations \cite{Tenberg2015} would enable further improvement. Reducing one noise source, either charge or hyperfine noise, generally also allows making gates less sensitive to the other noise source since optimal gates will exploit tradeoffs between the sensitivity to different types of decohering noise.

Our results also indicate that the unavoidable presence of nuclear spins in GaAs spin qubits, which is often thought of as prohibitive for their technological prospects, actually does not preclude the fidelities required for fault-tolerant quantum computing. This could allow leveraging other strengths of GaAs compared to Si, such as a small effective mass leading to relaxed fabrication requirements, the absence of near-degenerate valleys and a direct band gap potentially enabling optical interfacing. Although driven by the needs of GaAs based \stz qubits, we expect that our approach is equally viable for other encoded spin qubits facing similar difficulties, and can be adapted for implementing exchange-mediated two-qubit gates.  

\section{Methods}
\subsection{Qubit system}
We work in a dilution refrigerator at an electron temperature of about \SI{130}{mK} using the same sample as Ref. \citenumber{Botzem2015}. A lateral double quantum dot is defined in the two-dimensional electron gas of a doped, molecular-beam epitaxy-grown GaAs/AlGaAs-heterostructure by applying voltages to metallic surface gates. We use the same gate layout as Ref. \citenumber{Shulman2012} shown in \rfig{fig:main_1}b with two dedicated RF gates (yellow) for controlling the detuning. As we only apply RF pulses to these gates and no DC bias, we can perform all qubit operations without the need for bias tees, which reduces pulse distortions.

Quantum gates are performed in the (1,1) charge configuration, where one electron is confined in the left and one in the right quantum dot. In this regime, the computational subspace is defined by the $s_z = 0$ triplet state \qT and the spin singlet state \qS. The other $s_z = \pm 1$ (1,1) triplet states $\ket{\uparrow\uparrow}$ and $\ket{\downarrow\downarrow}$ are split off energetically via the Zeeman effect by applying an external magnetic field of \SI{500}{mT}.

We always readout and initialize the dot in the $\{\qUD, \qDU\}$ basis by pulsing slowly from (0,2) to (1,1) and thus adiabatically mapping singlet \qS and triplet \qT to \qUD and \qDU (see supplementary information).

\subsection{Readout calibration}
For measuring the quantum state, we discriminate between singlet and triplet states by Pauli spin blockade. Using spin to charge conversion \cite{Petta2005}, the resistance of an adjacent sensing dot depends on the spin state and can be determined by RF-reflectometry. In this manner, we obtain different readout voltages for singlet and triplet states but cannot distinguish between $\ket{\mathrm{T}_0}$ and the triplet states $\ket{\mathrm{T}_{\pm}}$.

The measured voltages are processed in two ways. First, binning on the order of $10^4$ consecutive single shot measurements yields bimodal histograms where the two peak voltages roughly correspond to the singlet and triplet state. Second, the measured voltages are averaged over many repetitions of a pulse to reduce noise.

For self-consistent state tomography, we linearly convert the averaged voltages to probabilities $p(\ket{0})$. The parameters of the linear transformation are obtained by fitting the histograms \cite{Barthel2009} with a model that takes the decay and excitation of \qS and \qT during the readout phase into account.

Due to the long gate sequences, the benchmarking experiments in \rfig{fig:main_2} are expected to produce a sizable leakage state population $\ket{\mathrm{T_+}}$. We have attempted to include $\ket{\mathrm{T_+}}$ explicitly in the histogram fit model but found that this introduces too many additional parameters. In order to achieve an approximate calibration, we prepare and readout a completely mixed state once in about $10^3$ measurements as an additional reference. We then enforce $p(\ket{0}) = 0.5$ for the mixed state voltage $U_{\mathrm{M}}$ in the histogram fits. While we have not quantitatively analyzed the error from this approximate procedure, we suspect that the suboptimal contrast in \rfig{fig:main_2} might be related. 

For GAMBIT, the averaged voltages $U_i$ corresponding to the error syndromes $S_i$ do not need to be explicitly converted to probabilities $p(\ket{0})$. Since mixed and triplet state reference voltages $U_{\mathrm{M}}$ and $U_{\mathrm{T}}$ are measured alongside the error syndromes, it is attractive to directly minimize $\tilde{U_i} = |U_i - U_{\mathrm{M}}|$ for $i = 1 \ldots 6$ and $\tilde{U_i} = |U_i - U_{\mathrm{T}}|$ for $i \in \{7, 8 \}$. Adjusting the contrast of $\tilde{U_i}$ with the aid of histograms can improve convergence and yields the expressions for $\tilde{S_i}$ from the main text.

Note that GAMBIT, RB and all other fits used in this work are insensitive to state preparation and measurement (SPAM) errors. Therefore, our readout calibration does not need to be especially accurate or precise. The only figure which is sensitive to SPAM errors is the singlet probability of the asymptote in \rfig{fig:main_2}a. However, the measured value $p(\qZ) = \SI{0.36}{}$ deviates significantly from $0.5$ so that leakage is the most plausible explanation for the observed second decay rate, irrespective of SPAM errors.

Further information regarding readout can be found in the supplementary information.


\subsection{Acknowledgments}
This work was supported by the Alfried Krupp von Bohlen und Halbach Foundation, DFG grant BL 1197/2-1, BL 1197/4-1, SFB 689 and the Deutsche Telekom Foundation.

\subsection{Author contributions}
D.B. and D.S. carried out molecular-beam-epitaxy growth of the sample used in this work. T.B. fabricated the sample and set-up the experiment. S.S.H. developed the driver for the digitizer hardware used for data acquisition. T.B. and P.C. developed the feedback software and conducted the experiment. H.B., T.B. and P.C. analyzed the data and co-wrote the paper. H.B. and P.C. developed the theoretical models.

\subsection{Additional information}
The authors declare no competing financial interests. Online supplementary information accompanies this paper. Correspondence and requests for materials should be addressed to H.B.


\bibliographystyle{naturemag_noURL_noarxiv}
\bibliography{bibliography/ex_hifi_gates_paper}
\end{document}